\documentclass[10pt]{article}

\usepackage{setspace}
\usepackage{amsmath}
\usepackage{amssymb}
\usepackage{amsfonts}
\usepackage{graphicx}
\usepackage[utf8]{inputenc}
\usepackage{slashed}
\usepackage{physics}
\usepackage{mathrsfs}
\usepackage{bbm}	
\usepackage{cancel}	
\usepackage[dvipsnames]{xcolor}
\usepackage{float}      
\usepackage{cite}

\usepackage[margin=1in]{geometry}  
\graphicspath{{figures/}}

\newcommand{\nn}{\\ \nonumber}
 
\newcommand{\pd}{\partial}

\begin{document}

\begin{center}

    {\large \bf Entanglement and local holography in quantum gravity}

    \vskip .75cm

\text{Gabriel Wong}

    \vskip .5cm

\text{Oxford Math Institute, Radcliffe Observatory}\\

\text{Andrew Wiles Building, Woodstock road, Oxford}\\

\text{OX2 6GG, United Kingdom}

    \vskip .75cm
\textit{gabrielwon@gmail.com}
\end{center}

    \vskip .75cm


\begin{abstract}
\noindent 
The It from Qubit paradigm proposes that gravitational spacetimes emerge from quantum entanglement.   So far, the main evidence for this involves holographic dualities, where the entangled qubits live in a dual nongravitational theory.    In this essay, we argue that string theory provides the mechanism to define these entangled qubits in the bulk gravitating theory.   This involves a local form of geometric transition, which is the stringy mechanism that underlies \emph{local} holography.  We illustrate how this works in the A model topological string,
\end{abstract}

\vspace{4.0cm}

\begin{center}

\text{Essay written for the Gravity Research Foundation}

\text{2025   Awards for Essays on Gravitation}

\text{Submission date: March 31,2025}

\end{center}

\doublespacing
\section{Introduction } 
Over the past decade, tremendous progress has been made in understanding the emergence of gravity from the entanglement structure of a dual quantum theory.  This is exemplified in AdS/CFT duality, where geometric features of the semi-classical bulk are related to entanglement measures in a dual CFT living at asymptotic infinity.   A key element in this dictionary is the quantum extremal surface (QES) formula\cite{Ryu:2006bv, Engelhardt:2014gca} , relating gravitational entropy, defined in terms of areas of bulk extremal surfaces, to boundary entanglement entropy.  However, very little is known about how spacetime emerges directly from the bulk quantum theory: this is important because that is where observers like ourselves reside !   The intuition from AdS/CFT suggests that the entanglement entropy of the bulk string theory should play a key role.  However, there are immediate obstacles to defining bulk entanglement measures, which require a division of the gravitational degrees of freedom into subsystems.   In QFT, given an appropriate regulator, the degrees of freedom can be split into spatial subregions.   Quantum mechanically, this means that given a Cauchy slice $\Sigma$ on a fixed background, and a division $\Sigma = V \cup \bar{V} $ into subregions, one can define a  factorization map 
\begin{align}\label{fact1}
    i_{\epsilon}: \mathcal{H}_{\Sigma} \hookrightarrow \mathcal{H}_{V} \otimes \mathcal{H}_{\bar{V} },
\end{align} where $\epsilon$ is a suitable regulator.   But what would such a splitting mean in the bulk gravity theory in which spacetime is fluctuating?  Indeed a spatial subregion is not a diffeomorphism invariant concept.  At the quantum level, the presence of gravitational constraints would seem to forbide a factorization  into independent subregion Hilbert space.  Thus, we are led to the question: 
\\
\\
   \emph{Is there a generalization of the factorization map \eqref{fact1} in quantum gravity, leading to an entanglement entropy that reproduces the area of extremal surfaces in the low energy limit? } 
\\

In this essay, we will argue that string theory provides the mechanism for such a factorization. 
The main idea is to interpret closed strings moving in a gravitational background as entangled open strings in a background with a large N number of branes/antibranes.  These \emph{entanglement} branes play the role of the entangling surface in QFT, separating subsystems labelled by $V$ and $\bar{V}$.
Note that, unlike in QFT, the ``subsystems" live in a different background than the global one.    This is a key feature of string theory, in which branes can dissolve into a closed string background via a geometric transition.    We will show that this phenomenon can be leveraged to factorize the closed string Hilbert space, and to define a trace on the open string Hilbert space which can be used to compute entanglement entropy \cite{Donnelly:2020teo,Jiang:2020cqo}.     

Since geometric transitions provide the stringy mechanism for holography, our construction is closely related to the latter.  In fact, the type of geometric transition relevant for defining entanglement entropy in string theory is simply a local version of holography.    To explain this fact, and to connect our stringy entanglement entropy to gravitational entropy, we will first review the extended Hilbert space framework for defining entanglement entropy in QFT.   In particular, we formulate this construction via the Euclidean path integral, and explain the notion of a ``shrinkable" entanglement boundary condition \cite{Donnelly:2018ppr,Jafferis:2019wkd}.  After proposing a quantum gravity analogue of the shrinkable boundary, we will give an illustrative example in the A model topological string, where the shrinkablility arises from the geometric transition of D branes. 

\section{Local holography and the ``shrinkable'' boundary}
\paragraph{The QFT  story} In  relativistic QFT, the Hilbert space $\mathcal{H}_{\Sigma}$ on a cauchy slice $\Sigma$ does not factorize into subregions: 
$$\mathcal{H}_{\Sigma} \neq   \mathcal{H}_{V}\otimes  \mathcal{H}_{\bar{V} }
$$ 
This is because the  tensor product on the RHS includes states of arbitrarily high energy due to  arbitrarily singular behavior of the quantum field at the entangling surface.   On the other hand, the physical Hilbert space $\mathcal{H}_{\Sigma}$ only contains finite energy excitations above the global vacuum.    This is an artifact of the contiuum and can be addressed by introducing a regulator $\epsilon$ separating $V$ and its complement. This  produces a ``stretched" co dimension 1 entangling surface $S_{\epsilon}$. 

$$ \includegraphics[scale=.16]{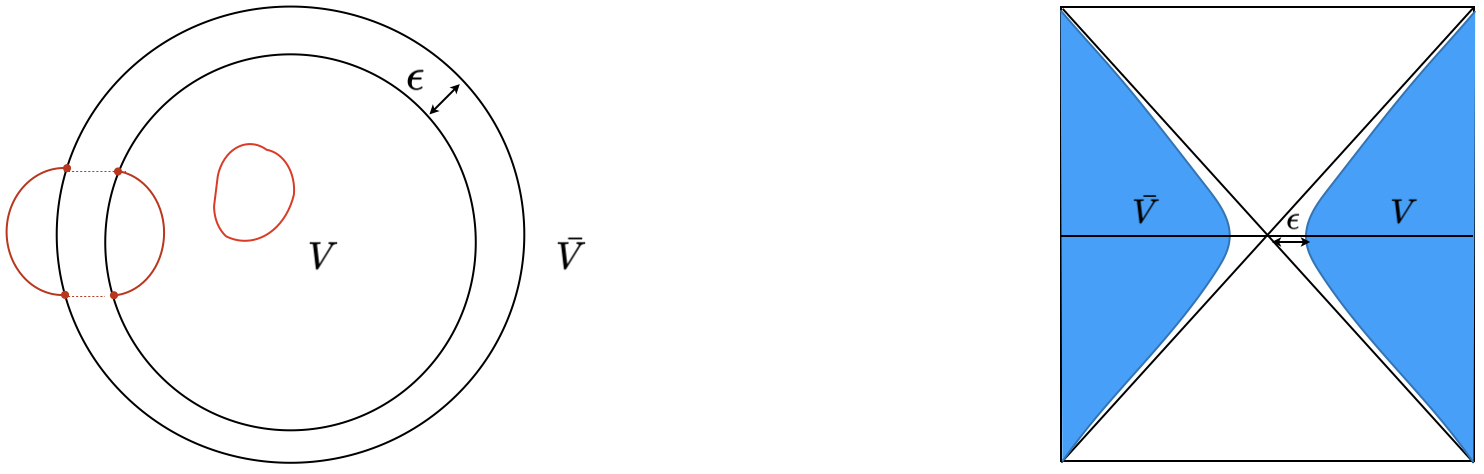} $$

However, in gauge theories, there is a further subtlety due to gauge constraints\footnote{As depicted in the left figure, these constraints arise from Wilson loops that cross the entangling surface} which cross the entangling surface: these are present even on the lattice, creating a tension with Hilbert space factorization.  
 This problem is solved by lifting the gauge constraints that cross the entangling surface: this liberates would-be pure gauge modes near $S_{\epsilon}$ and extends the subsystem Hilbert space $\mathcal{H}_{V}$ to include these \emph{edge modes} degrees of freedom.   The entanglement of these edge modes is crucial for gluing together the subregions and contributes to the entanglement entropy of the subregion \cite{Donnelly:2011hn,Donnelly:2014gva}.    

Since gravity is a gauge theory, it is natural to apply the same formalism.   To highlight the analogy with holography, we will formulate the extended Hilbert space construction in terms of the Euclidean path integral.  Thus, we interpret the  factorization map  $i_{\epsilon}$ in \eqref{fact1} as a Euclidean process the cuts open the Cauchy slice \cite{Donnelly:2018ppr,Jafferis:2019wkd}.   Because the entangling surface is not a physical boundary, we must ensure that the associated boundary condition  does not change the correlations of the original,  unfactorized state.  This is achieved by imposing the isometric property on $i_{\epsilon}$, shown in the left figure below.
\begin{align}\label{shrink}
\vcenter{\hbox{
\includegraphics[scale=.15]{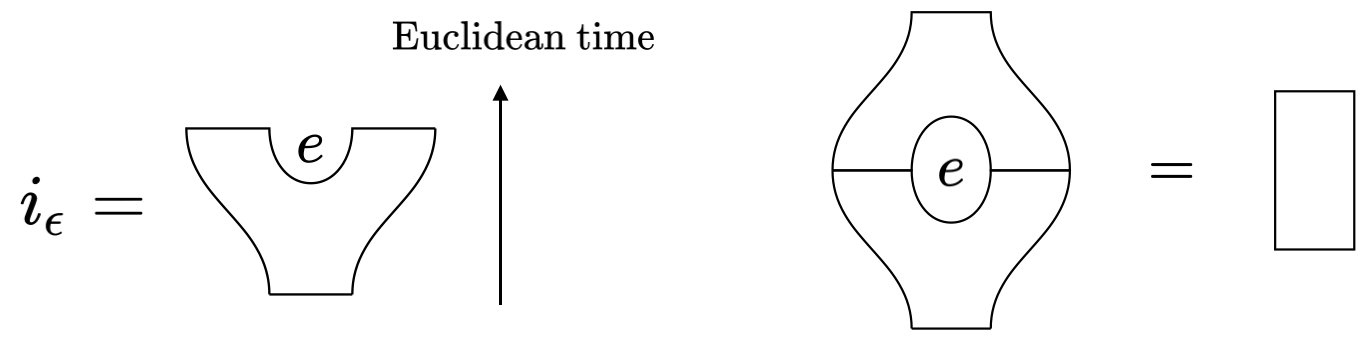}}}
\qquad \qquad 
\vcenter{\hbox{\includegraphics[scale=.15]{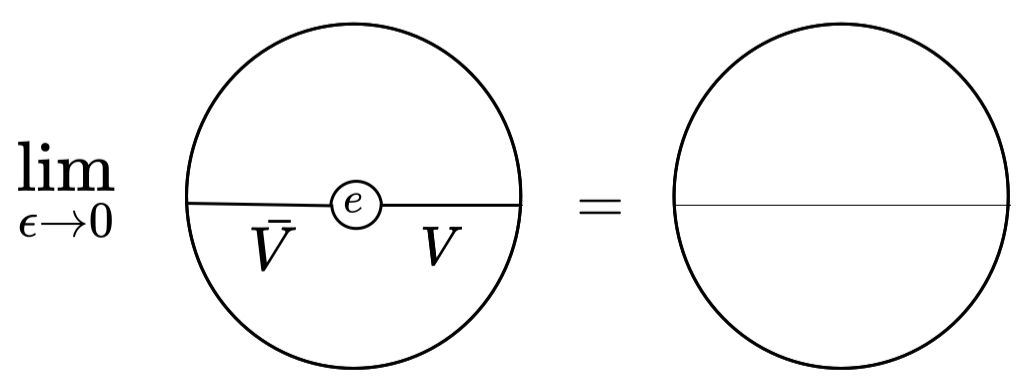}}}
\end{align}

On the right figure, we derived an example of a shrinkability condition by applying the isometry condition to a ``Hartle-Hawking" state prepared by a path integral on a half disk.
This ensures that the path integral on the full disk produces a trace on the subregion Hilbert space, so 
\begin{align} \label{Zdisk}
    Z_{\text{disk}} =  \tr_{V} \rho_{V},
\end{align}
where $\rho_{V}$ is the reduced density matrix on the subregion V.  This is because the the lower half of the annulus prepares a tensor product state. Then gluing along $\bar{V}$ is a partial trace that produces $\rho_{V}$, while the gluing along $V$ computes $\tr_{V} \rho_{V}$.   In QFT, the trace over $V$ has to be  renormalized  due to UV divergences from the shrinking limit $\epsilon \to 0$.  This is the same divergence that appears in Euclidean Rindler space, associated to the infinite temperature of the Rindler horizon.  

Figure \eqref{shrink} suggests a type of local holography, where degrees of freedom in an infinitestmal Euclidean disk is captured by edge degrees of freedom at its boundary.   Indeed, one can show that in abelian gauge theories \cite{Ball:2024hqe} and linearized gravity \cite{Anninos:2020hfj,Law:2025ktz}, shrinkability \eqref{Zdisk} fails in the absence of the edge modes.  However, in QFT this type of holography is approximate in the sense that an infinite subtraction is needed to relate the disk and the annulus. 

Another consequence of shrinkability is that the Von Neumann entropy defined by $S_{\text{Vn}}=-\tr_{V} \rho_{V}\log \rho_{V}$ 
 coincides with geometric entropy, defined as the response of a path integral to the insertion of a conical singularity of strength $2\pi - \beta$
\begin{align}
S_{\text{geo}} = (1-\beta \pd_{\beta})|_{2\pi} \log Z_{\text{disk}}(\beta).
\end{align}
Here $\beta$ parametrizes the angular range around the center of the disk.  This is the QFT analogue of the Gibbons Hawking entropy in gravity.   The matching of $S_{\text{Vn}}=S_{\text{geo}}$ depends delicately on the entanglement entropy of the edge modes, and has been worked out in detail in abelian gauge theory\cite{Donnelly:2014fua,Donnelly:2015hxa,Ball:2024hqe}. 
\paragraph{The gravity story}
In QFT, the shrinking limit produces a UV divergence in the entanglement entropy $S_{\text{Vn}}$:  this matches the divergence in $S_{\text{geo}}$, which is due to the insertion of the conical singularity.    On the other hand, geometric entropy in gravity is defined by the Gibbons-Hawking prescription \cite{Gibbons1977}, which gives a \emph{finite} entropy by incorporating the fluctuations of spacetime.

Rather than inserting a conical singularity into the bulk, one varies the size $\beta$ of an asymptotic boundary circle, while allowing the bulk geometry to back react  to a \emph{smooth }solution to Einstein's equations.  In this context, the disk should be viewed as a cigar geometry appearing in a cross section of the Euclidean black hole, and the Gibbons-Hawking prescription gives the gravitational entropy  $S= \frac{A}{4G}$.  

The finiteness of the Gibbons-Hawking entropy suggests that quantum gravity provides a more sophisticated way to 
cut off the spacetime near the tip of the cigar.  

Indeed, the size of the Euclidean circle behaves as a local inverse temperature, which approaches zero at the tip.  As the circle approaches the Planck length, we should allow the possibility of a dual description where the tip is replaced by quantum gravity degrees of freedom that effectively cuts off the space time.    Such a holographic duality relating gravity on the annulus and the cigar would naturally provide a  bulk state counting interpretation to the Gibbons Hawking entropy.  This is a local generalization of the more standard holographic dictionary where the gravitating disk is dual to  a \emph{non-gravitational} theory at its asymptotic boundary.

In the gravity version of the shrinkability condition, the analogue of the QFT edge modes living at the rigid cutoff surface $S_{\epsilon}$ is given by dynamical  objects which ``end the spacetime" \cite{Mathur:2014nja,McNamara:2022xkg}.   To fully appreciate this viewpoint, consider the Lorentzian interpretation of the shrinkablility condition \eqref{shrink}. This is obtained by cutting the Euclidean geometries along the time reflection symmetric slice, then evolving in Lorentzian time \cite{Jafferis:2021ywg,Mertens:2022ujr}: 
\begin{align}\label{bulkEREPR}
\includegraphics[scale=.2]{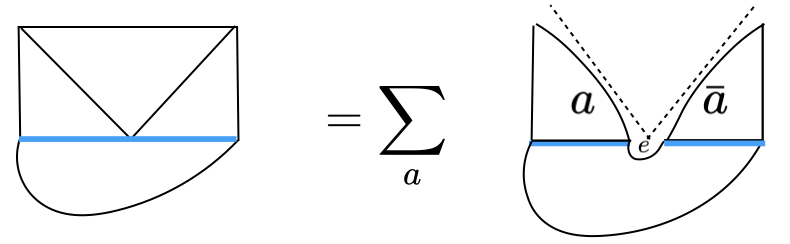}
\end{align}

The unfactorized Hartle Hawking state on the left evolves into a Lorentzian two sided black hole, whereas the factorized Hartle Hawking state  evolves into an entangled sum over singled-sided geometries.  This is the  bulk version of the equivalence \cite{Maldacena:2001kr} between the two-sided black hole and the boundary thermal field double state\footnote{The bulk version of this duality was used in the arguments of  \cite{VanRaamsdonk:2010pw} to discuss the emergence of the ER bridge from entanglement. However, so far there has been no explicit realization of this bulk version of ``ER=EPR" }, which now goes under the slogan ER=EPR \cite{Maldacena:2013xja}.  In its bulk incarnation, this equivalence requires a ``complete set" of gravitational objects that ends the single-sided spacetimes.  Note that these need not be actual boundaries of the spacetime manifold: for example, the spacetime can end by capping off an internal manifold not shown in the cross section above, as in well known examples of Fuzzball geometries\cite{Mathur:2013bra}.  The existence of such spacetime ending defects -complete with respect to the shrinkability condition- is a highly nontrivial statement.   However, there are indications that such ``cobordism defects" must exist in order for string theory to be compatible with the basic principle that no global symmetries exist in quantum gravity.  This is referred to as the cobordism conjecture \cite{McNamara:2019rup},  and our discussion show that this is intimately related to shrinkablility criteria and the equivalence between entanglement entropy and gravitational entropy\footnote{ The connection between the cobordism conjecture and ER=EPR was first discussed in  \cite{McNamara:2022xkg} }.   In the context of Hilbert space factorization in string theory, we refer to these cobordism defects generically as  ``entanglement branes".   We expect these objects to be the microscopic constituents that make  up the unfactorized gravitational background.   Below, we will consider an example where these constituents are branes.
\section{Hilbert space factorization and entanglement entropy in topological string theory}
\paragraph{Shrinkable boundary condition and geometric transitions}
 The A model topological string propagates on  six dimensional target spaces.  However, we will restrict ourselves to backgrounds where the string amplitudes are described by a 2D theory\cite{Aganagic:2004js,Bryan:2004iq}. These backgrounds are Calabi Yau manifolds that are the direct sum of 2 vector bundles over a Riemann surface $\Sigma$. Such a bundle is represented  by the surface $\Sigma$ and two integers $k_{1},k_{2}$ that label the Chern classes of the bundles: 
$$ 
\includegraphics[width=0.4\linewidth]{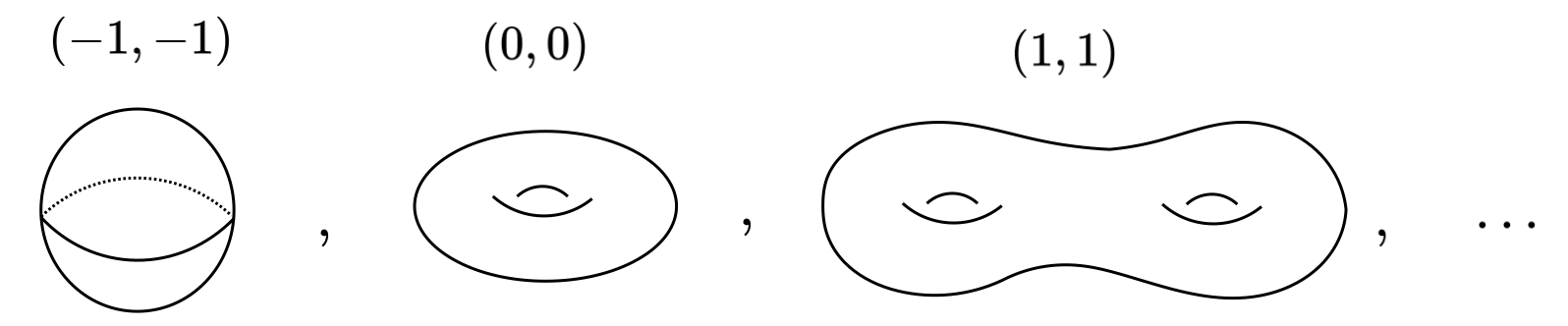}
$$
 The Calabi Yau condition requires  $k_{1}+k_{2}= -\chi (\Sigma)$ , where $\chi(\Sigma) $ is the Euler Characterstic : this ensures that the spacetime is on shell so that Einstein's equations are satisfied.  
 
The full closed string amplitude on these geometries involve a sum over worldsheets\footnote{These are actually worldsheet instantons, which are holomorphic maps from the worldsheet to the target space} that wrap the surface $\Sigma$ with arbitrary winding number. Remarkably, these amplitudes are captured to all orders in the string coupling $g_{s}$ by the large N limit of $U(N)$, q-deformed 2D Yang Mills on $\Sigma$. To match the string amplitudes, the $q$ deformation parameter must be equal to $q=e^{ig_{s}}$.  Given the Chern class labels, the only other dependent parameter of the amplitudes is the complexified area $t$ of $\Sigma$, with the imaginary part capturing the B field flux.

Let's consider the closed string background corresponding to $\Sigma=S^2$.  This is the resolved conifold, which will be the analogue of the Euclidean black hole geometry with the base $S^2$ playing the role of the cigar.  
We define a geometric entropy by varying the closed string amplitude with respect to the size of a Euclidean circle on $S^2$, which produces conical singularities at the two antipodal points where this circle shrinks \cite{Donnelly:2020teo}.   Just like the Gibbons Hawking prescription, this is an \emph{on shell} variation because we preserve the Calabi Yau condition: we keep the bundle structure $(-1,-1)$ fixed in the variation so it matches the constant Euler characteristic of the base sphere under a conical variation. Because the amplitudes only depend on the area of $\Sigma$, this is the same as varying the area $t$ of the sphere $\Sigma$: 
\begin{align}\label{Sent}
S_{\text{geo}}= (1- t \pd_{t}) \log Z_{\text{res}}(t)
\end{align}

We will obtain a statistical interpretation of this quantity by  introducing two shrinkable holes in $S^2$, so that it becomes a cylinder describing a trace over a bulk Hilbert space.  This validity of this type of path integral logic in string theory is highly nontrivial: it works because the closed string amplitudes satisfy the same cutting and gluing rules as the path integral of a 2D theory,provided that we keep track of the Chern numbers $(k_{1},k_{2})$ which add component-wise as we glue.

The simplest example of such a cutting and gluing procedure is the decomposition of the sphere into two discs: 
$$
\vcenter{\hbox{\includegraphics[scale=.2]{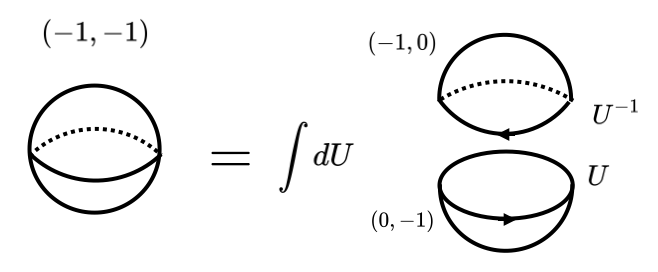}}}$$
  The disks contain a ``gluing" boundary, where we must specify  a boundary condition.  In the q- deformed 2DYM theory, these are labeled by the holonomies $U =\text{P} \exp \oint A$  of a $U(N)$ gauge field, in the limit  $N\to \infty$. These holonomies label states $\ket{U}$ that form a basis for the q2DYM Hilbert space on a circle, which we identify as the closed string Hilbert space.  The wavefunction of a single string winding n times around the circle is $\langle U|n\rangle=\tr U^n$, while multi string wavefunctions correspond to arbitrary products of these traces.  They span the Hilbert space
 $$ \mathcal{H}_{\text{closed} } = \text{Class functions on } U(\infty) $$
In string theory, each boundary of $\Sigma$ corresponds to  a large $N$ stack of branes or antibranes, and the boundary condition $U$ is the worldvolume holonomy.   These \emph{non compact} branes are 3 dimensional in the full spacetime and \footnote{They wrap Lagrangian manifolds on which the symplectic form vanishes} and have topology $ S^1 \times  \mathbb{R}^2$.  In the presence of these branes, the worldsheet develop boundaries that wind around the $S^1$ factor.  The gluing of the 2d manifolds corresponds to annhilation of brane-antibrane pairs that glue together the open string amplitudes \cite{Aganagic:2003db}.    Notice that these branes  are not co-dimension 1 objects, and therefore they do not cut the target space into separate pieces. However they do  give a well defined cutting  of closed string amplitudes into open string amplitudes and vice versa. 

We define the Hartle Hawking state as a wavefunctional\footnote{The fact that D brane amplitudes transform as wavefunctionals was shown in \cite{Aganagic:2003qj}} of the worldvolume gauge field $U$, obtained by cutting the resolved conifold partition function:
\begin{align}
\ket{HH} = \vcenter{\hbox{\includegraphics[scale=.08]{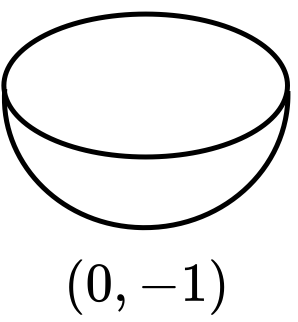}}},\qquad \qquad \qquad \vcenter{\hbox{\includegraphics[scale=.08]{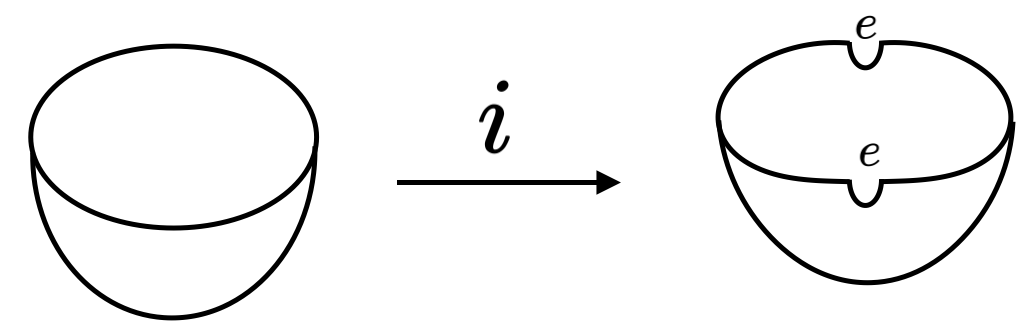}}}
\end{align}
This state is a condensate of winding strings, which we want to factorize into open strings as shown in the right figure.  This requires a shrinkable boundary condition, which was obtained in \cite{Donnelly:2020teo,Wong2025}:
\begin{align} \label{localH}
 \vcenter{\hbox{\includegraphics[scale=.1]{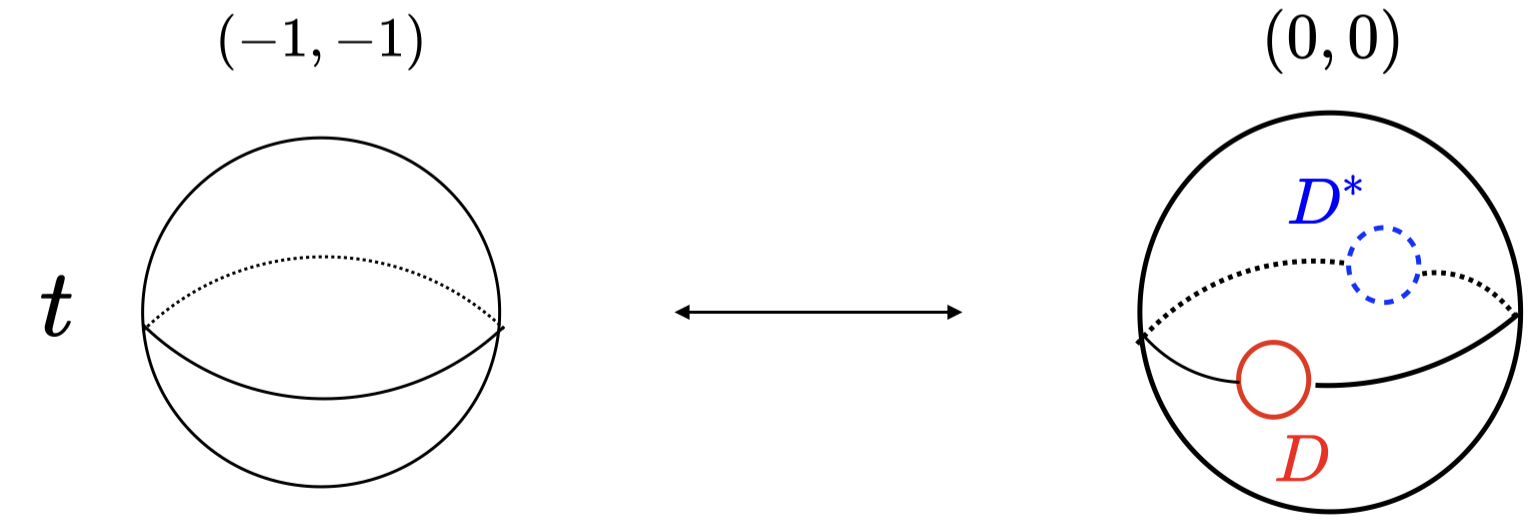}}}\qquad \qquad  D_{ij} = \delta_{ij} e^{ig_{s} (-j+1/2+N/2)}, \,\,t= ig_{s}N= \text{fixed as }N\to\infty.
\end{align} 
Here, the holonomy $D$ is a distinguished element of the quantum group $U(N)_{q}$ called the Drinfeld element.   In the string theory interpretation, \eqref{localH} describes a duality between closed strings on the resolved conifold and open strings on $\mathbb{C}^3$, with  a stack of branes and anti branes at special values $D$ and $D^*$ of the worldvolume holonomy\footnote{ In the figure, the change in the background is indicated by the change in the Chern numbers, which reflects the change in the Euler characteristic of $\Sigma$.}.  This duality holds in the large N limit, where we fix $t=i g_{s}N $; it is a generalized version of a geometric transition, in which  a large N number of branes and anti branes  dissolved into a closed string background with flux given by $t$.  

Indeed, to relate to the more conventional type of geometric transitions, where a single stack of branes is replaced by flux,  we can simply reduce to a single shrinkable boundary :
$\vcenter{\hbox{\includegraphics[scale=.065]{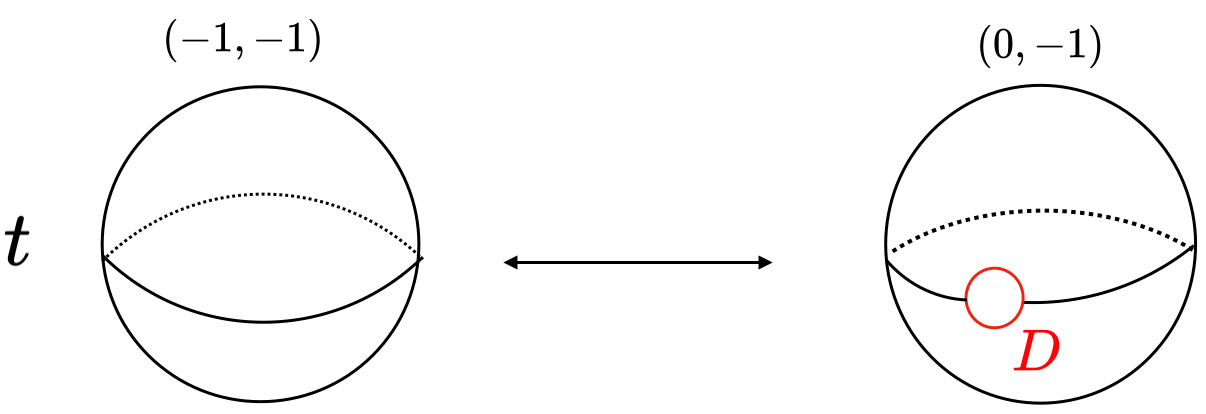}}}$\cite{Wong2025} .  The transition of these non compact branes in the A model were studied in\cite{Gomis:2006mv,Gomis:2007kz}, where they were characterized as  surgery operations in spacetime.   This involves removing a neighborhood of the branes with topology  $\mathbb{B}^3 \times S^{1} \times \mathbb{R}^2 $, and gluing in $S^2\times D^2 \times \mathbb{R}^2$, where $\mathbb{B}^3$ is a 3 ball surrounding the branes, and $D^2$ is the disk obtained by filling in the $S^1$ factor.  This is an analogue of the surgery operations which characterize the Euclidean AdS/CFT correspondence in type IIB string theory or M theory.
\paragraph{Hilbert space factorization and the quantum trace} 
In the open string background on $\mathbb{C}^3$, the A model amplitudes only involve worldsheets whose boundaries wind nontrivially around the $S^1$ on the branes,  which play the role of a putative thermal circle:
$$ \includegraphics[scale=.12]{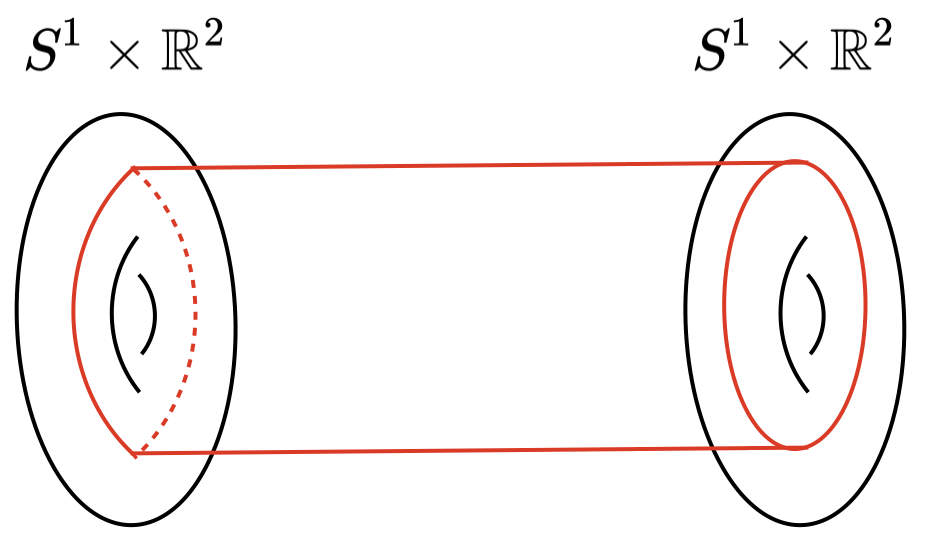}
$$

The thermal interpretation is manifest when $q=1$, corresponding to zero string coupling.  This is because the world volume holonomy $D$ reduces to the identity, which corresponds to a  boundary  condition $A=0$ that is local in the thermal time coordinate.  Locality in time allows us to interpret the worldsheet path integral as the trace of a open string Hamiltonian $H_{\text{open}}$ acting on a ``subregion" Hilbert space $L^{2}(U(\infty))$, obtained by quantizing open strings ending on these D branes\cite{Donnelly:2016jet}.

A basis for $L^{2}(U(\infty))$ is given by arbitrary products 
$\{ U_{i_{1}j_{1}} \cdots U_{i_{n}j_{n} }\}$ of matrix element of $U(N)$. 
Each matrix element is an open string, and $(i_{1},j_{1}), \cdots (i_{n},j_{n})$ are the Chan Paton factors labelling the different D branes.  On the other hand, the shrinkable boundary condition for $q\neq 1$ is non local in time, since it gives a nontrivial holonomy $D$ in the thermal circle.  However, due to the magic of quantum groups, this non local boundary condition still accomodates a trace interpretation, provided that we q deform the open string Hilbert space to:
$$\mathcal{H}_{\text{open}} = L^{2}(U(\infty)_{q} ) 
$$
This is spanned by a basis $U_{i_{1}j_{1}} \cdots U_{i_{n}j_{n}} $of quantum group matrix elements that do not commute. Remarkably, there is a quantum trace $\tr_{q}$ on $L^{2}(U(\infty)_{q} )$ satisfying the shrinkability condition:
$$
    Z_{res}(t) = \tr_{q} e^{ -t H_{\text{opem}}}\qquad \qquad   \vcenter{\hbox{\includegraphics[scale=.05]{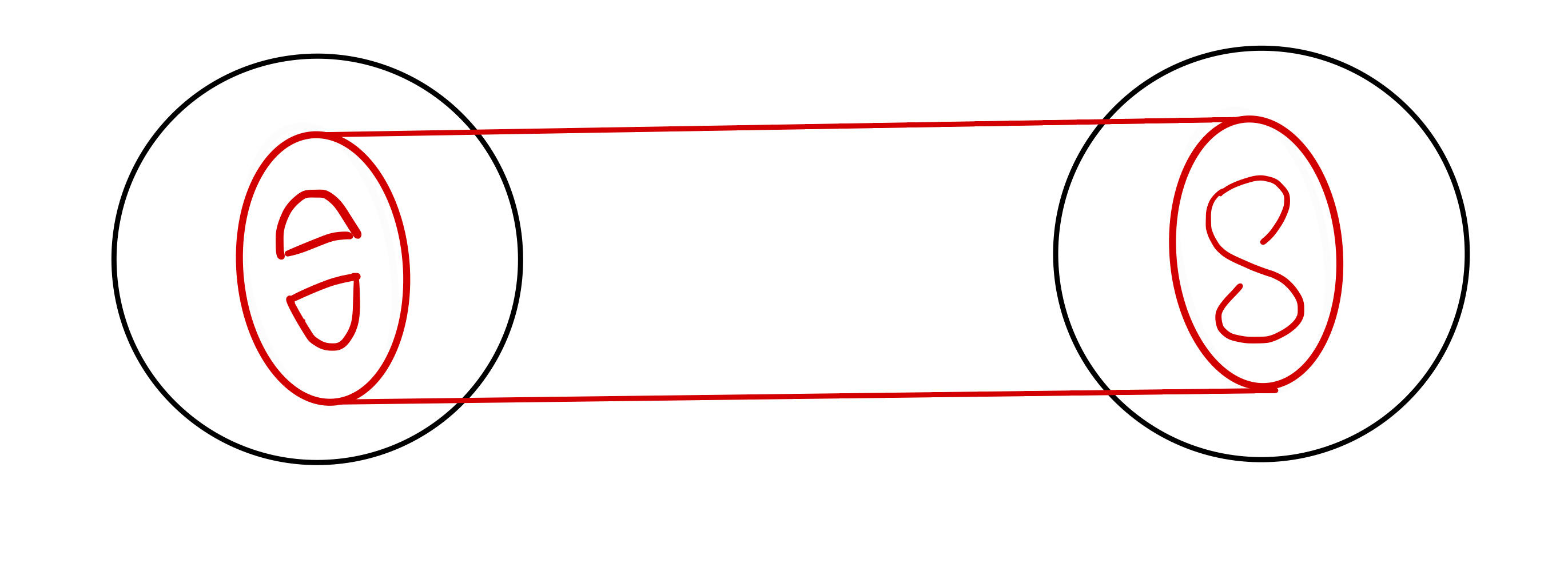}}}
$$
The quantum trace has a beautiful worldsheet interpretation \cite{Wong2025}, illustrated in the right figure.  The shrinkable boundary condition effectively compactifies\footnote{ There is a subtle difference with between this entanglement brane configuration and actually changing the background to include branes wrapping two $S^3$. In the latter case, there would be ribbon graphs of the Chern Simons theory that do not attach to the worldsheets, and these give extra contributions that are absent in the entanglememt brane configuration \cite{Jiang:2020cqo} }the entanglement branes into $S^3$, which supports a dynamical Chern Simons gauge field \cite{Gomis:2007kz,Wong2025}.  The end points of the open strings interact with this gauge field via ribbon graphs that attach themselves to worldsheet boundaries.  Summing these interactions to all orders deforms the ordinary trace into a quantum trace  \cite{Jiang:2020cqo}.  Thus,  $Z_{res}(t)$ is counting the dressed Chan Paton factors of the open strings.

The same trace is used to define the factorization map cutting closed strings to open strings:
\begin{align*}
i: \mathcal{H}_{\text{closed}}  &\rightarrow \mathcal{H}^{1}_{\text{open}} \otimes \mathcal{H}^{2}_{\text{open}}\nn
 \tr(U^n) &\rightarrow \tr_{q}(U_{1} U_{2})
\end{align*}
 This map takes the closed string  $U=P\exp \oint A $, and cuts it into open strings $U_{1} =P \exp \int_{a}^{b} A$ and $U_{2}= Pexp\int_{b}^{a}  A $ by contracting the Chan Paton indices appropriately using the quantum trace. 
In the string theory language, the factorization introduces two additional set of large N brane-antibrane pairs that cut the closed string loops in $\mathcal{H}_{closed} $ into entangled open strings.  Thus, the factorization is implemented by an intersecting configuration of branes in $\mathbb{C}^3$ \cite{Wong2025}.

Finally, the quantum trace can be employed to define a Von Neumann entropy $S_{q}=- \tr_{q} \rho \log \rho $ that measures the entanglement of the Chan Paton factors, and equals the geometric entropy \eqref{Sent} of the resolved conifold. 
While $\tr_{q}$ satisfy more exotic properties than the usual Hilbert space trace, it defines a good measure of entanglement that has been applied to study entanglement entropies of anyons and $q$ deformed spin chains \cite{Bonderson:2017osr,Couvreur_2017}.  
\section{Closing thoughts}
We have argued that string theory provides the microscopic degrees of freedom needed to factorize the quantum gravity Hilbert space and define entanglement entropy in a bulk gravitational theory.   We illustrated this in topological string theory, where closed strings are entangled open strings, and the stringy edge modes are Chan Paton factors labeling D branes.  It maybe possible to bring this example even closer to the geometric ER=EPR picture of \eqref{bulkEREPR} by considering the representation basis for the closed and open string Hilbert spaces.  Then each closed string state  corresponds to a Wilson loop in a particular representation $R$ of $U(N)$, and these states are known to have a geometric description in terms of bubbling Calabi Yau geometries\cite{Gomis:2006mv,Gomis:2007kz}.  This suggests that open string states obtained from cutting these Wilson loops may correspond to bubbling Calabi Yau's with added features captured by the Chan Paton factors.   This would give a fully backreacted version of ER=EPR. 

De Sitter quantum gravity is a natural arena in which these ideas might find their application. Unlike in AdS, dS has no spatial boundary, so gravity is never turned off.  Nevertheless, Gibbons Hawking showed that the de Sitter horizon carries entropy, and it is tempting to interpret this as the entanglement entropy of bulk quantum gravity\footnote{Recently, an argument was proposed which interpreted de Sitter  entropy of Chan Paton factors for spacetime filling D branes\cite{Dvali:2024dzf}. 
}.  Indeed, from the perspective of the q2DYM theory, our stringy entanglement entropy is computing the de Sitter entropy of a 2D Hartle Hawking state.  

Finally, we note that in 2 and 3 dimensions, it is possible to apply the ``edge mode" paradigm directly in the low energy gravity theory, in which the entangling surface is allowed to fluctuate \cite{Blommaert:2018iqz,Mertens:2022ujr,Wong:2022eiu}.    Here, the shrinkable boundary condition that matches the Gibbons Hawking prescription is necessarily non local in Euclidean time \cite{Jafferis:2019wkd}.  Nevertheless, just like in the topological string, one can define a notion of a subsystem Hilbert space as well as a trace that satisfies the shrinkability condition.  The bulk entanglement entropy then has an edge mode contribution. In a saddle point approximation, it gives 
$$  S_{\text{edge}} \sim \log \dim R^*= \frac{A}{4G}$$
where $R^*$ is the saddle point representation of the edge mode symmetry.  Remarkably, this formula reproduces the area of black holes horizons and certain extremal surfaces in the QES formula \cite{McGough:2013gka,Blommaert:2018iqz,Mertens:2022ujr,Wong:2022eiu}. This begs the question of whether we can relate these low energy edge modes to more microscopic, stringy degrees of freedom. 
To address this question, it maybe useful to first consider a gauge theory analogue  given by Maxwell theory in 4D.   Here, the edge mode entanglement entropy in pure Maxwell theory matches the entanglement entropy  due to  the effects of charged particles \cite{Donnelly:2015hxa,Ball:2024hqe,Casini:2019nmu}, so it is reasonable to suspect a relation between the two.

\section*{Acknowledgements}
It is a pleasure to thank Jan de Boer,  William Donnelly, Dan Jafferis, Jake Mcnamara, and Takuya Okuda for useful discussions. GW is supported by STFC grant
ST/X000761/1, the Oxford Mathematical Institute, and Harvard CMSA.

\bibliographystyle{utphys}
\bibliography{topstring}

\end{document}